# Synthesis, crystal and magnetic structure of iron selenide $BaFe_2Se_3$ with possible superconductivity at $T_c=11K$

A. Krzton-Maziopa[1], E. Pomjakushina[1], V. Pomjakushin[2], D. Sheptyakov[2], D. Chernyshov[3], V. Svitlyk[3] and K. Conder[1]

[1]*Laboratory for Developments and Methods, Paul Scherrer Institute, 5232 Villigen PSI, Switzerland*
[2]*Laboratory for Neutron Scattering, Paul Scherrer Institute, 5232 Villigen PSI, Switzerland*
[3]*Swiss-Norwegian Beam Lines at ESRF, BP220, F-38043 Grenoble, France*

E-mail: kazimierz.conder@psi.ch

**Abstract.** We report on synthesis of single crystals of $BaFe_2Se_3$ and study of their crystal and magnetic structures by means of synchrotron single crystal X-ray and neutron powder diffraction. The crystal structure has orthorhombic symmetry and consists of double chains of $FeSe_4$ edge connected tetrahedra intercalated by barium. Below 240 K long range block-spin checkerboard antiferromagnetic (AFM) order is developed. The magnetic structure is similar to one observed in $A_{0.8}Fe_{1.6}Se_2$ (A=K, Rb or Cs) superconductors. The crystals exhibit a transition to the diamagnetic state with an onset transition temperature of $T_c$ ~11 K. Though we observe FeSe as an impurity phase (<0.8% mass fraction) the diamagnetism unlikely can be attributed to the FeSe-superconductor which has $T_c \approx 8.5K$.

## 1. Introduction

With the discovery of alkaline metal intercalated FeSe superconductor [1] with transition temperature $T_C \approx 30K$ a chalcogenide analog to 122 iron pnictides superconductors [2,3] was found. Shortly after the first reported synthesis [1] of $K_{0.8}Fe_2Se_2$ (K122) further compounds with a formula $A_xFe_{2-y}Se_2$ intercalated with A=Cs [4], Rb [5], Tl/K [6] and Tl/Rb [7] were successfully prepared. For all these superconductors critical temperatures around 30K were found. All the intercalated chalcogenides, opposite to pnictides, can be relatively easily grown as high quality single crystals by Bridgman method. This is extremely important as for many studies e.g. neutron scattering experiments $cm^3$ size single crystals are requested. For $Cs_{0.8}Fe_{2-y}Se_2$ and $K_{0.8}Fe_{2-y}Se_2$ it was found that the empirical dependence of $T_C$ on so known anion height (the distance between Fe and the chalcogenide/pnictide layers) as stated for pnictides [8] is also fulfilled [4]. Thus, further increase of $T_C$ (maximum value for pnictides ~55K) could be potentially possible by applying pressure. Unfortunately, negative pressure effect with suppression of superconductivity above 8 GPa for $Cs_{0.8}Fe_{2-y}Se_2$ was observed.[9] Further for the $A_xFe_{2-y}Se_2$ a coexistence of superconductivity and a strong magnetism ($T_N \approx 478.5K$) was found [10]. Based of neutron scattering experiments structure models revealing iron vacancy and magnetic orderings were proposed [11, 12, 13].

In our attempt to explore ternary iron chalcogenides we took attention to alkali earth metal families. According to Iglesias and Steinfink [14] in the case of barium transition metal chalcogenides $Ba_xM_yX_z$ (where M=3d metal, X=Se or S) compounds with different stoichiometries x:y:z of 2:1:3, 1:1:2, 1:4:3, 1:2:2 and 1:2:3 could be synthesized. For all these compounds nine different structure types have been found all of them based on columns of face-sharing $BaX_6$ triangular prisms (tetrahedra). In $BaMX_2$ (Ba112) and $BaM_2X_2$ (Ba122) sheets are created by edge sharing of the neighboring columns whereas both in $BaM_2X_3$ (Ba123) and $BaM_4X_3$ (Ba143) the columns are isolated [14]. Especially Ba123 seems to be interesting analog to K122 as additional charge of the alkali earth metal is balanced by additional chalcogenide anion in the formula. Figure 1 shows crystal structure model of orthorhombic Ba123 emphasizing coordination of the Fe atoms. In contrast to alkali metal intercalated chalcogenides (like K122) the Fe-Se layers are here not two-dimensional but consist of double chains of $FeSe_4$ edge connected tetrahedra.

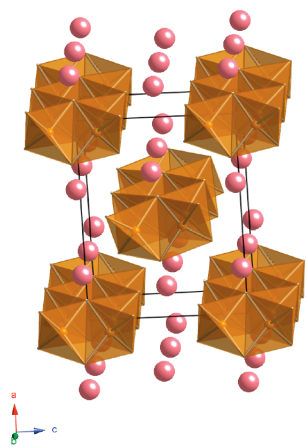

**Figure 1.** Crystal structure of Ba123 selenide. In this representation coordination polyhedra for Fe are shown instead of those for Ba. The double chains of Fe-Se tetrahedrons are edge shared and propagated along b-axis.

In this work $BaFe_2Se_3$ (Ba123) single crystals were grown for the first time by the Bridgman method. Opposite to the K122 single crystals Ba123 are stoichiometric showing no vacancies both in alkali earth and iron sublattices. Further iron-selenium layer in the structure has $Fe_2Se_3$ stoichiometry and is created by double chains of $FeSe_4$ edge connected tetrahedra. Ba123 is showing long antiferromagnetic order at temperatures below 250K and traces of superconductivity with onset of $T_C \approx 11$.

## 2. Experimental

Single crystals of BaFe$_2$Se$_3$ were grown from the melt using the Bridgman method. High purity (at least 99.99%, Alfa) powders of iron and selenium were mixed in the proper proportion and pressed into rod. For the single crystal synthesis a piece of Fe$_2$Se$_3$ rod was sealed in a double-wall evacuated silica ampoule with a stoichiometric amount of pure barium. The ampoule was annealed at 1150$^{\circ}$C for 24 h for homogenization. Afterwards the melt was cooled down to 750$^{\circ}$C with a rate of 6$^{\circ}$C/h and kept at this temperature for the next 20h. Finally the furnace was cooled down to room temperature at the rate 200$^{\circ}$C/h. Well-formed black crystal rods of 7 mm diameter (inside diameter of the quartz ampoule) were obtained. The crystals could be easily cleaved into plates with flat shiny surfaces. The phase purity of the grown crystals was characterized by powder x-ray diffraction (XRD) using D8 Advance Bruker AXS diffractometer with Cu K$\alpha$ radiation. For these measurements a fraction of the crystal was cleaved, powderized, and loaded into the low background airtight specimen holder. All the work was performed in a He glove box to protect the powder from an oxidation.

Homogeneity and elemental composition of the cleaved crystals were studied using x-ray fluorescence spectroscopy (XRF, Orbis Micro-XRF Analyzer, EDAX). Elemental distribution maps for Ba, Fe and Se were collected in vacuum applying white X-ray radiation produced by Rh-tube (35kV and 500μA). The x-ray primary beam was focused to a spot of 30 μm diameter. Primary beam Ti filter (25μm thickness) was applied. An area of ~0.5 cm$^2$ was scanned. Prior to the measurements elemental calibration was done using as a standard carefully weighted, homogenized and pressed into pellet mixture of Se, Fe and BaCO$_3$. The applied calibration procedure results in ~1.5% accuracy of the determined stoichiometric coefficients values.

Single crystal X-ray data for BaFe$_2$Se$_3$ were collected on the SNBL beamline BM1A at the ESRF synchrotron in Grenoble (France) with the KUMA6 diffractometer equipped with the TITAN CCD detector, $\lambda$ = 0.69736(1)Å, using a combination of ω-and φ-scans of 1º and a readout pixel resolution of 62 $\mu$m. The sample was mounted in a cryo-loop and cooled down to 150 K with the Oxford Cryostream 700+ N$_2$ blower. Distance from crystal to detector was calibrated using lattice dimensions of the powder sample prepared from the crushed crystals. Intensities were integrated with CrysAlis [15]. Empirical absorption correction was made with SADABS procedure [16]. Structure solution was performed with SHELXS and the structure refinement with SHELXL97 [17]. Powder diffraction data were collected at the MAR345 detector at the same beamline as the single crystal (SNBL, ESRF). Crystal-to-detector distance was calibrated with LaB$_6$ NIST standard, the data were processed with Fit2D [18] and local software was used to estimate standard deviations of the measured intensities.

Neutron powder diffraction experiments (NPD) were carried out at the SINQ spallation source of the Paul Scherrer Institute (PSI, Switzerland) using the high-resolution diffractometer for thermal neutrons, HRPT [19], with the neutron wavelength $\lambda$=1.886Å. The sample was loaded into a vanadium container with an indium seal in a He glove box. Refinement of crystal and magnetic structures from powder diffraction data were carried out with FULLPROF program [20], with a use of its internal tables for scattering lengths and magnetic form factors. The ac magnetization measurements were performed on a quantum design PPMS magnetometer.

## 3. Results and discussion

Micro-XRF analysis performed on the $BaFe_2Se_3$ crystals confirmed their homogeneity. Stoichiometric coefficients determined on the basis of micro-XRF mapping were: 0.99±0.02; 1.99±0.04 and 3.00±0.05 for Ba, Fe and Se respectively.

Figure 2 shows three slices of the reciprocal space for Ba123 single crystal obtained by synchrotron diffraction. One can see a pronounced broadening of reflections that agrees with needle-like shape of highly fragile crystals. Such a broadening is also seen in the reciprocal maps normal to the *a\** direction, while the *h0l* section shows much narrower profiles. This observation is consistent with the morphology of the single crystals, which are assembled from numerous blocks misaligned by a rotation presumably around the longest *a*-axis. We have also noted a weak rod-like diffuse scattering propagating along the *a\** that could be tentatively attributed to the stacking faults normal to the *a* direction. For many tested crystals the diffraction data can be successfully indexed assuming few twin components rotated for few degrees relative to each other. Crystal morphology, mosaicity and diffuse signal agree with quasi 1D character of the crystal structure that is built from covalently bound double chains of $FeSe_4$ tetrahedra weakly packed together by Ba ions.

The results of single crystal structural refinement of $BaFe_2Se_3$ are summarized in Tables 1 and 2. The structural model proposed from single crystal experiment well fit the powder x-ray and neutron diffraction data (Figure 2). Notably, there is one very weak additional reflection at d≈5.52Å at T=300K seen only in x-ray powder diffraction pattern that is not modeled with the structure of $BaFe_2Se_3$ (*Pnma* space group). This peak can be tentatively assigned to the $FeSe_{1.974}$ with the lattice constants a=3.7738, c=5.5235Å [11]. Refinement of this impurity phase with the structure model from [21] gives 0.8% mass fraction. At variance with the powder data, a summation of the observed frames in single crystal experiment does not evidence an additional reflection at about 5.52 Å; such a procedure implies an averaging of the single crystal data set over all orientations of the same crystal and is similar to a powder average over many crystals.

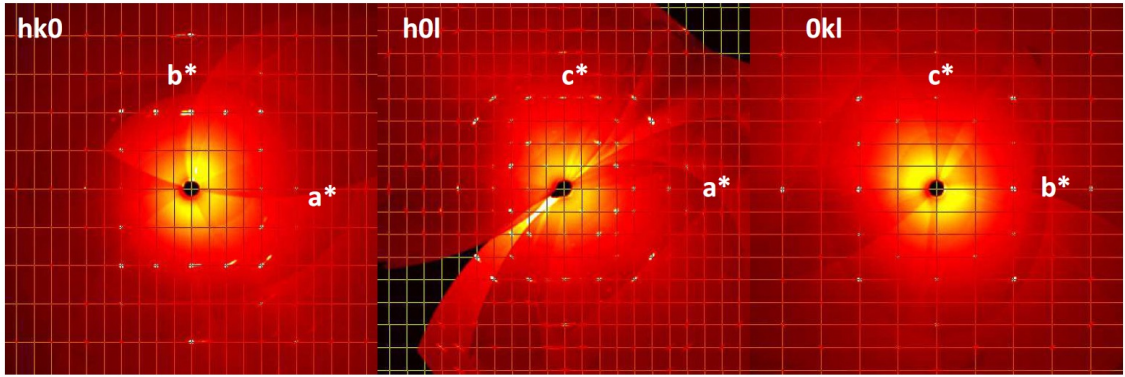

**Figure 2.** Three slices of the reciprocal space for Ba123 single crystal obtained by synchrotron diffraction at T=150 K.

On cooling below T=240K, a set of additional Bragg peaks is observed in the neutron powder diffraction patterns that can be indexed with the propagation vector k = [1/2,1/2,1/2]. The *Pnma* space group with k=[1/2,1/2,1/2] has two two-dimensional irreducible representations (irreps) that both enter 6 times the reducible magnetic representation for the Fe spin sitting in the general (8d) position. We have sorted all possible symmetry adapted solutions with a restriction that the Fe spins have the same magnitude and are oriented either parallel or antiparallel to each other and being directed along crystal axes. There are two solutions for both $\tau_1$ and $\tau_2$ irreps that excellently fit the experimental NPD pattern as shown in Fig. 3 with only one refinable parameter – the moment value. The fit reliability R-factors for these two models are the same in the refinements of powder diffraction data. All other magnetic models give significantly worse R-factors (30-60%). The Fe spins have moment of 2.1$\mu_B$ and directed along a-axis, i.e. perpendicular to the $Fe_3Se_2$ double chains. The projection of the magnetic structure on the (bc)-plane is shown in Figure 4. Interestingly that the magnetic configuration $\tau_1$ with the four spins forming ferromagnetic block is similar to the block-spin checkerboard structure observed in $A_{0.8}Fe_{1.6}Se_2$ (A=K, Rb or Cs) [12]. The description of the structure model and the magnetic and structure parameters and the details of the refinements at 2K are summarized in Table 3.

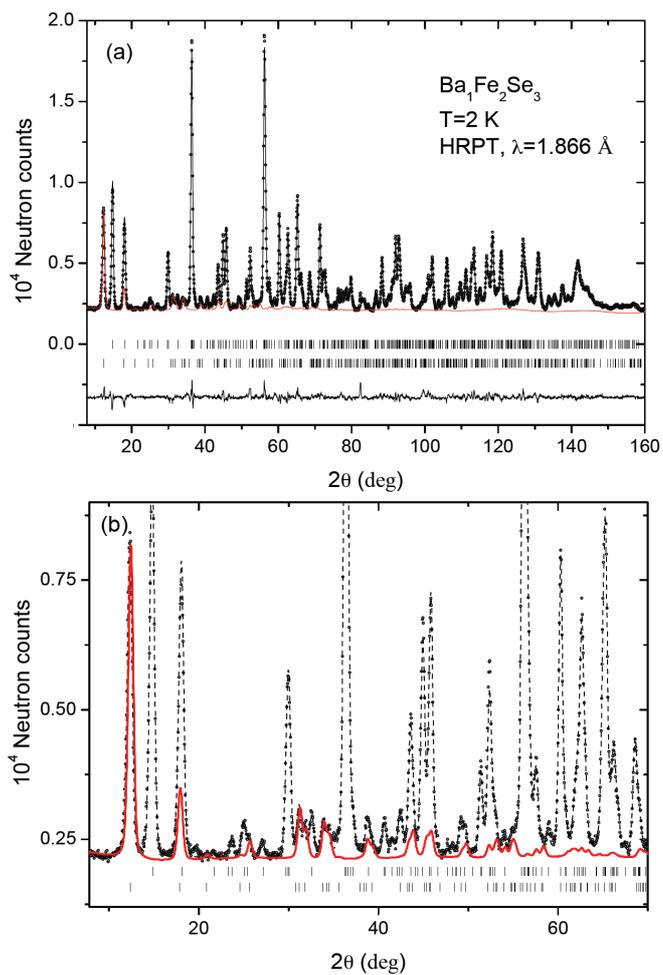

**Figure 3.** (a) Neutron powder diffraction pattern, the calculated profile and difference plot for BaFe$_2$Se$_3$ at 2K. The rows of ticks show the Bragg peak positions for the nuclear and magnetic phases. The magnetic contribution together with the background is shown by the red line. (b) Fragment of the diffraction pattern at low 2Θ showing the magnetic contribution.

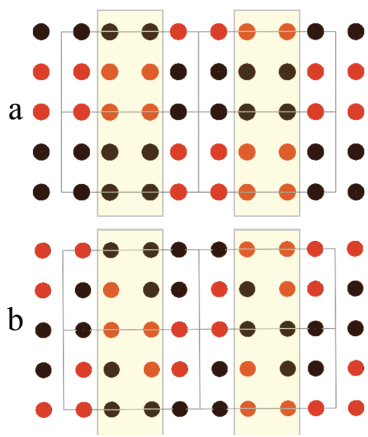

**Figure 4.** Refined magnetic structure models (a) and (b) for Ba123. The magnetic structures for $\tau_1$ (a) and $\tau_2$ (b) are shown in the projection on the (bc) plane. The black and red circles correspond to the up and down spin directions. The $Se_2Fe_3$ double chains separated by approximately a/2 distance along a-axis are shown on different background color. See the erratum for the Figure 4b after the References section at the end of the document.

Figure 5 shows a temperature dependence of the magnetic moment on Fe. It is evident that the long range AFM disappears quite abruptly above 240K. However the short range AFM correlations persist up to room temperature. At 240K the magnetic peak width starts to increase and at 250K and higher the peak width is about 10 times larger than at low temperature (see Figure 5b). The domain sized estimated from the AFM Bragg peak broadening amounted to >650Å at low temperature and about 20Å at room temperature.

a.

b.

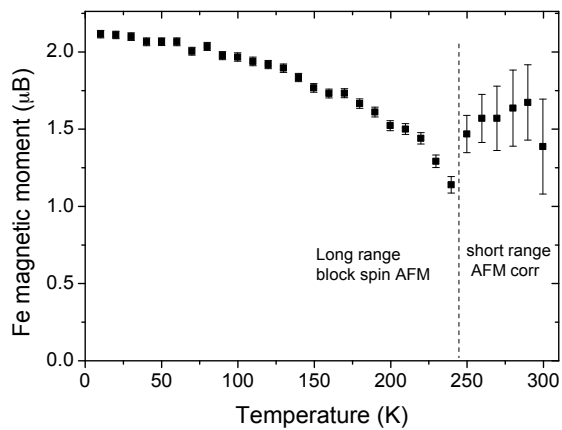
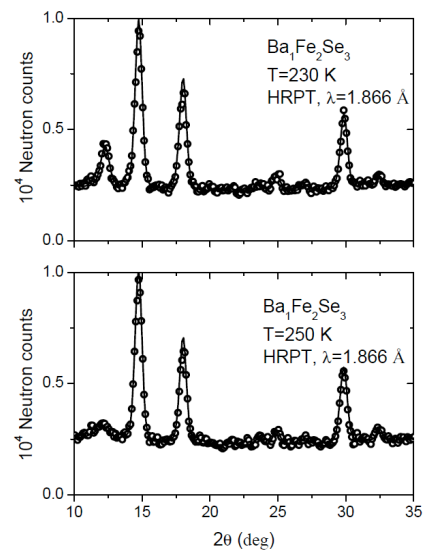

**Figure 5.** a) Temperature dependence of the magnetic moment on Fe. Above $T_N$=240K the long range AFM order is destroyed, and the moments in this region correspond to the fit of the broad Lorentian peak centered approximately at zero satellite (1/2,1/2,1/2) at $2\Theta \approx 12°$ assuming the same magnetic model; b) NPD patterns taken at T=230 and 250K. Note the change of the magnetic peak width at $2\Theta \approx 12°$.

Figure 6 presents a temperature dependence of the real part of ac magnetic susceptibility for BaFe$_2$Se$_3$ sample measured in magnetic field of 1 Oe and frequency 1000 Hz. The sample exhibits a clear transition to the diamagnetic, possibly superconducting state with an onset transition temperature above T$_c$~11K. However, the possible superconducting volume fraction is only about 1%. For comparison a typical magnetization curve for FeSe$_{0.98}$, the only stated impurity phase, is also shown in the same plot. One can clearly see that the onset transition for BaFe$_2$Se$_3$ is at least 2.5K higher than those for FeSe$_{0.98}$ (8.33(11) [21], 8.21(4) [22]). Therefore, the diamagnetism above 8.5K can not be attributed to the presence of FeSe impurity phase.

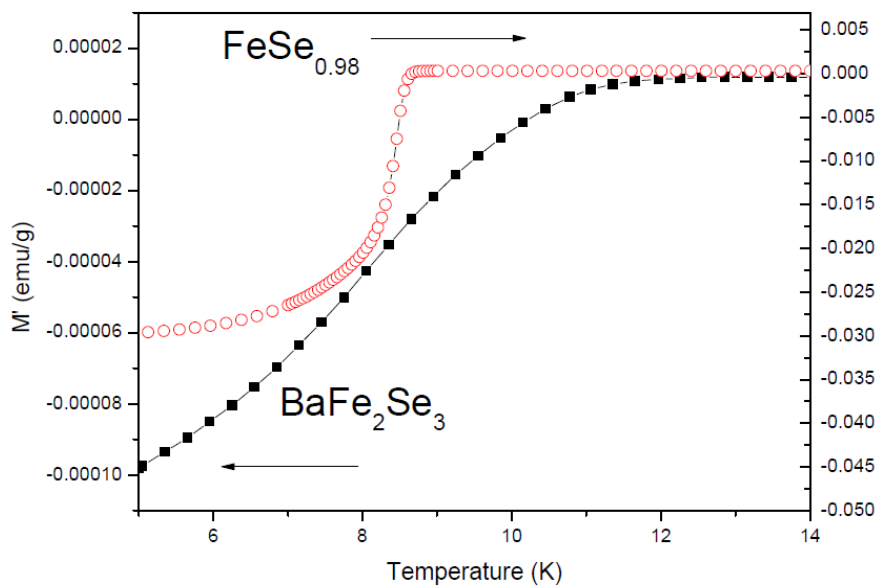

**Figure 6.** Temperature dependence of the real part of the ac magnetic susceptibility for BaFe$_2$Se$_3$ (black squares) together with FeSe$_{0.98}$ (red circles).

## 4. Summary

Successful crystal growth of alkali earth intercalated iron selenide BaFe$_2$Se$_3$ (Ba123) is reported. The crystal structure is similar to analog alkali metal intercalated chalcogenides with the difference being one dimensional arrangement of FeSe$_4$ edge connected tetrahedra creating double chains running along b-axis within Fe$_2$Se$_3$ bc-layers. The AFM long range order is observed below T$_N$≈240K. The antiferromagnetic structure contains 4-spin ferromagnetic blocks oriented antiparallel along the Fe$_2$Se$_3$ chains with Fe moments of 2.1μ$_B$ along a-axis. Short range AFM is preserved at least up to room temperature. Diamagnetic transition with an onset at about 11K suggests a possible existence of a superconducting state.

**Table 1.** Atomic coordinates (x $10^4$) and equivalent isotropic displacement parameters ($Å^2$ x $10^3$) for 150K. U(eq) is defined as one third of the trace of the orthogonalized $U^{ij}$ tensor.

|       | x       | y    | z       | U(eq) |
|-------|---------|------|---------|-------|
| Ba(1) | 1854(1) | 2500 | 5223(1) | 16(1) |
| Se(2) | 3569(2) | 2500 | 2308(2) | 13(1) |
| Se(3) | 6285(2) | 2500 | 4902(2) | 14(1) |
| Se(4) | 4007(2) | 2500 | 8158(2) | 16(1) |
| Fe(5) | 4937(2) | 7(3) | 3526(2) | 13(1) |

**Table 2.** Anisotropic displacement parameters ($Å^2$ x $10^3$) for 150K. The anisotropic displacement factor exponent takes the form: $-2\pi^2[h^2 a^{*2} U^{11} + ... + 2hk a^* b^* U^{12}]$

|       | $U^{11}$ | $U^{22}$ | $U^{33}$ | $U^{23}$ | $U^{13}$ | $U^{12}$ |
|-------|----------|----------|----------|----------|----------|----------|
| Ba(1) | 24(1)    | 6(1)     | 18(1)    | 0        | 3(1)     | 0        |
| Se(2) | 21(1)    | 4(1)     | 15(1)    | 0        | -3(1)    | 0        |
| Se(3) | 23(1)    | 5(1)     | 13(1)    | 0        | -1(1)    | 0        |
| Se(4) | 23(1)    | 6(1)     | 18(1)    | 0        | 5(1)     | 0        |
| Fe(5) | 17(1)    | 8(1)     | 12(1)    | 0(1)     | 0(1)     | 0(1)     |

**Table 3.** Crystal and magnetic structure parameters of $BaFe_2Se_3$ refined from NPD at 2K in *Pnma* space group. The isotropic atomic displacement parameter B is in $Å^2$. $\chi^2$ is the global chi-square (Bragg contribution), $R_B$ is a Reliability Bragg factors are given for the nuclear and magnetic phases respectively, m is the magnetic moment amplitude in $\mu_B$ - all are given for two magnetic models ($\tau_1$, $\tau_2$).

| a (Å) | 11.89033(14) |
|-------|--------------|
| b (Å) | 5.40798(5)   |

| | x | y | z | B | Mult. |
|---|---|---|---|---|---|
| c (Å) | 9.13752(11) | | | | |
| Ba1 | 0.18514(30) | 1/4 | 0.52208(51) | 0.442(72) | 4 |
| Fe1 | 0.49404(15) | 0.00148(38) | 0.35263(18) | 0.695(25) | 8 |
| Se1 | 0.35503(25) | 1/4 | 0.23284(32) | 0.777(56) | 4 |
| Se2 | 0.63036(21) | 1/4 | 0.49047(31) | 0.434(50) | 4 |
| Se3 | 0.40149(23) | 1/4 | 0.81731(29) | 0.398(47) | 4 |
| | m∥a Fe1 | $\chi^2$ | | $R_{B(nucl)}$ | $R_{B(magn)}$ |
| $\tau_1$ | 2.101 (13) | 3.82 | | 3.04 | 4.76 |
| $\tau_2$ | 2.101 (13) | 3.82 | | 3.03 | 4.84 |

**Acknowledgements**

The authors thank the Sciex-NMS$^{ch}$ (Project Code 10.048), the Swiss National Science Foundation and NCCR MaNEP for the support of this study. This study was partly performed at Swiss neutron spallation SINQ of Paul Scherrer Institute PSI (Villigen, PSI).

**References**

[1] Guo J, Jin S, Wang G, Wang S, Zhu K, Zhou T, He M and Chen X 2010 *Phys. Rev.* B **82** 180520(R)
[2] Rotter M, Tegel Mand Johrendt D 2008 *Phys Rev Let.* **101** 107006
[3] Sasmal K, Lv B, Lorenz B, Guloy A M, Chen F, Xue Y-Y, and Chu C-W 2008 *Phys Rev Let.* **101** 107007
[4] Krzton-Maziopa A, Shermadini Z, Pomjakushina E, Pomjakushin V, Bendele M, Amato A, Khasanov R, Luetkens H and Conder K 2011 *J. Phys.: Condens. Matter* **23** 052203
[5] Wang A F, Ying J J, Yan Y J, Liu R H, Luo X G, Li Z Y, Wang X F, Zhang M, Ye G J, Cheng P, Xiang Z J, and Chen X H. 2011*Phys. Rev. B* **83** 060512
[6] Fang M, Wang H, Dong C, Li Z, Feng C, Chen J, and Yuan H Q 2011 *Europhys. Lett.* **94** 27009
[7] Hang-Dong Wang, Dong Chi-Heng, Li Zu-Juan, Mao Qian-Hui, Zhu Sha-Sha, Feng Chun-Mu, Yuan H Q, and Fang Ming-Hu 2011 *Europhys. Lett.***93** 47004
[8] Mizuguchi Y, Hara Y, Deguchi K, Tsuda S, Yamaguchi T, Takeda K, Kotegawa H, Tou H and Takano Y 2010 *Supercond. Sci. Technol.* **23** 054013
[9] Seyfarth G, Jaccard D, Pedrazzini P, Krzton-Maziopa A, Pomjakushina E, Conder K, Shermadini Z 2011 *Solid State Comm.* **151** 747
[10] Shermadini Z, Krzton-Maziopa A, Bendele M, Khasanov R, Luetkens H, Conder K, Pomjakushina E, Weyeneth S, Pomjakushin V, Bossen O, Amato A 2011 *Phys. Rev. Lett* **106** 117602
[11] Pomjakushin V Yu, Sheptyakov D V, Pomjakushina E V, Krzton-Maziopa A, Conder K, Chernyshov D, Svitlyk V, Shermadini Z 2011 *Phys. Rev. B* **83** 144410

[12] Pomjakushin V Yu, Pomjakushina E. V, Krzton-Maziopa A, Conder K, and Shermadini Z 2011 *J. Phys.: Condens. Matter* **23** 156003
[13] Bao W, Huang Q, Chen G F, Green M A, Wang D M, He J B, Wang X Q and Qiu Y 2011 *Chinese Phys. Lett.* **28** 086104
[14] Iglesias J E , Steinfink H 1975 *Z. Kristallogr* **142** 398
[15] CRYSALIS Software System, Ver. 171.34.44 (Oxford-diffraction Ltd., Oxford, England, 2011)
[16] Sheldrick G M 2002 *SADABS,Version 2.06* (University of Goettingen, Germany)
[17] Sheldrick G M 1997 *SHELXL97* (University of Goettingen, Germany)
[18] Hammersley A P, Svensson S O, Hanfland M, Fitch A N, Häusermann D 1996 Two-Dimensional Detector Software: From Real Detector to Idealised Image or Two-Theta Scan *High Pressure Research* **14** 235-248
[19] Fischer P, Frey G, Koch M, Könnecke M, Pomjakushin V, Schefer J, Thut R, Schlumpf N, Bürge R, Greuter U, Bondt S, and Berruyer E 2000 *Physica* B **276-278** 146
[20] Rodriguez-Carvajal J 1993 *Physica B* **192** 55
[21] Pomjakushina E, Conder K, Pomjakushin V, Bendele M, and Khasanov R 2009 *Phys. Rev.* B **80** 024517
[22] Khasanov R, Bendele1e M, Conder K, Keller H, Pomjakushina E, and Pomjakushin V 2010 *New Journal of Phyics.* **12** 073024

*Erratum*

# The synthesis, and crystal and magnetic structure of the iron selenide BaFe₂Se₃ with possible superconductivity at T<sub>c</sub> = 11 K

A Krzton-Maziopa[1], E Pomjakushina[1] V Pomjakushin[2], D Sheptyakov[2], D Chernyshov[3], V Svitlyk[3] and K Conder[1]

[1] Laboratory for Development and Methods, Paul Scherrer Institute, 5232 Villigen PSI, Switzerland
[2] Laboratory for Neutron Scattering, Paul Scherrer Institute, 5232 Villigen PSI, Switzerland
[3] Swiss–Norwegian Beam Lines at ESRF, BP220, F-38043 Grenoble, France

Due to technical mistake we have erroneously taken wrong computer file to plot the magnetic structure of the solution that we have found for the irreducible representation (irrep) τ2, which was shown in Fig. 4b. Table 3 showing also the details of magnetic structure refinements for τ2 and Fig. 4a were correct. The corrected Figure 4 is shown in this erratum. The structure (b) looks similar to (a) but the geometry of the Fe-plaquettes forming ferromagnetic 4-spins block is slightly different for (a) and (b) cases of different τ1 and τ2 symmetry. For completeness we explicitly list the symmetry operators, coordinates and spin directions +1 or -1 along a-axis for all eight Fe atoms in zeroth cell for τ1 (a) and τ2 (b) structures.

| Symmetry operator | x,y,z | Spin direction for a, b |
|---|---|---|
| x,y,z | 0.494, 0.001, 0.353 | +1, +1 |
| -x+1/2,-y,z+1/2 | 0.006, 0.999, 0.853 | -1, +1 |
| -x,y+1/2,-z | 0.506, 0.501, 0.647 | -1, +1 |
| x+1/2,-y+1/2,-z+1/2 | 0.994, 0.499, 0.147 | -1, -1 |
| -x,-y,-z | 0.506, 0.999, 0.647 | -1, -1 |
| x+1/2,y,-z+1/2 | 0.994, 0.001, 0.147 | +1, -1 |
| x,-y+1/2,z | 0.494, 0.499, 0.353 | -1, +1 |
| -x+1/2,y+1/2,z+1/2 | 0.006, 0.501, 0.853 | -1, -1 |

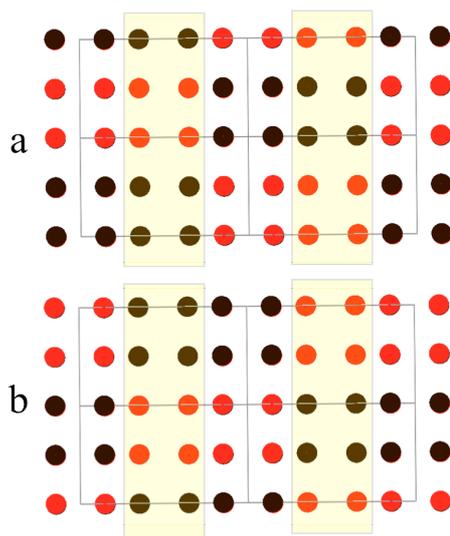

**Corrected Figure 4.**